# Highly Efficient Room Temperature Spin Injection in a Metal-Insulator-Semiconductor Light-Emitting Diode.

P. Van Dorpe[1*], V.F. Motsnyi[2], M. Nijboer[1], E. Goovaerts[2], V.I. Safarov[3], J. Das[1)], W. Van Roy[1], G. Borghs[1] and J. De Boeck[1]

*[1]Interuniversitary MicroElectronics Center (IMEC), Kapeldreef 75, B-3001 Leuven, Belgium*

*[2]University of Antwerpen-UIA, Universiteitsplein 1, B-2610 Antwerpen, Belgium*

*[3] Groupe de Physique des Etats Condensé (GPEC), Departement de Physique, Faculté des Sciences de Luminy, 13288 Marseille, France*

We demonstrate highly efficient spin injection at low and room temperature in an AlGaAs/GaAs semiconductor heterostructure from a CoFe/AlOx tunnel spin injector. We use a double-step oxide deposition for the fabrication of a pinhole-free AlOx tunnel barrier.  The measurements of the circular polarization of the electroluminescence in the Oblique Hanle Effect geometry reveal injected spin polarizations of at least 24% at 80K and 12% at room temperature.

---

*  Electronic mail: pvandorp@imec.be

Spintronic semiconductor-based devices that operate at room temperature require highly polarized currents inside the semiconductor[1]. Until recently, the injection of highly polarized currents was only demonstrated at low temperatures and high magnetic fields by the use of dilute magnetic semiconductors as spin injectors[2]. Ferromagnetic metals are very promising for room temperature spintronic applications due to their high Curie temperature, relatively large room temperature spin polarizations and the knowledge on their physical properties. However, there is a fundamental problem concerning the injection of spin-polarized currents from ferromagnetic metals in semiconductors. Schmidt *et al.*[3] showed that, due to a conductivity mismatch between the ferromagnetic metal and the semiconductor, diffusive transport through an ohmic contact does not allow for reasonable spin injection efficiencies. The inclusion of a large spin-dependent resistance between the ferromagnet and the semiconductor can circumvent this problem. Such a resistance can be provided by a tunnel barrier[4] as supported by recent calculations by Rashba[5] and Fert[6].

A few groups have successfully taken this approach. Zhu *et al.*[7] and Hanbicki *et al.*[8] reported an injected spin polarization of respectively 2% and 13% using a Schottky tunnel barrier between GaAs and epitaxially grown Fe. In the light emitting diode (LED) heterostructure the injected spin polarization was monitored through a polarization analysis of the light, created in a quantum well. Their geometry requires perpendicularly magnetized injector structures. For the in-plane Fe, a large magnetic field (2T) is required to saturate the thin film in the direction perpendicular to the film.

Recently[9,10] we have introduced the Oblique Hanle Effect (OHE) technique to overcome this problem. The first measurements on non-optimised spin-LED's already showed successful electrical spin injection into a semiconductor from a ferromagnetic metal at 80K. In this experiment the polarization of the injected electrons also was monitored through a polarization analysis of the emitted light in a surface emitting metal-insulator-semiconductor (MIS) spin-LED. However it deviates from the former in two ways. Firstly, our approach to spin injection is the incorporation of a thin amorphous $AlO_X$-layer as a tunnel barrier between the ferromagnetic metal and the semiconductor. (An $AlO_X$ tunnel barrier has also been used by Akinaga et al.[11]) Secondly, instead of changing the direction of the spin orientation inside of the ferromagnetic metal by saturating the magnetization out of the plane, we have manipulated the spin in the semiconductor, introducing the Oblique Hanle effect approach for spin injection assessment. The first measurements on non-optimised devices showed an injected spin polarization of at least 9 % at 80K.

This report focuses on the demonstration of highly efficient spin injection, persisting to room temperature in a MIS spin-LED. To improve the electrical properties of the tunnel oxide, we used a new multi-step Al deposition/oxidation process. This allows a better control over the oxidation process and produces devices with a higher optical efficiency. Using the devices with the improved tunnel barrier we demonstrate experimentally a high injected spin polarization of at least 24 % at 80K and persisting to 12 % at room temperature.

We have fabricated MIS spin-LEDs with the following structure: $p^+$ GaAs substrate / 200nm p-$Al_{0.3}Ga_{0.7}As$ ($2x10^{18}cm^{-3}$) /100 nm GaAs (undoped)/15 nm $Al_{0.2}Ga_{0.8}As$ (undoped) / 1.3 nm $AlO_x$/2nm CoFe/8 nm NiFe/ 5 nm Cu. The semiconductor heterostructure was grown by molecular beam epitaxy. Immediately after the growth, the sample was transferred into the sputtering machine for the fabrication of the tunnel injector. The oxide layer was deposited in a two-step process, which facilitates a full oxidation of the Al, reduces the chance on pinholes[12] and enables the fabrication of a thicker barrier. First, a 1nm thin Al layer is sputtered and naturally oxidized in a controlled oxygen atmosphere at 140. After sputtering and oxidizing of the first Al layer, a second Al layer was sputtered and oxidized, followed by the deposition the ferromagnetic stack. All metals are dc magnetron sputtered. The LED processing is described in more detail in ref[10]. In comparison with ref[10], the thickness of the $AlO_x$-layer is increased to allow a better optical efficiency due to the relative change of the tunnel probabilities for electrons and holes. This allows us to perform optical measurements at higher temperatures and at lower current stress applied to the tunnel oxide. Moreover recent calculations[5,6] suggest that enlarging the spin-dependent interface resistance induced by the tunnel barrier should result in an increase of the spin injection efficiency. The transmission electron microscopy (TEM) image of the MIS-structure (Fig. 1) shows that the tunnel-oxide is smooth, exhibits no pinholes and has atomically sharp interfaces with both the AlGaAs and the FM. A more extensive set of devices is under evaluation to verify the theoretical dependence of spin injection on barrier thickness.

Under forward bias spin polarized electrons are injected from the ferromagnetic top layer into the undoped 100 nm wide GaAs active region. There they radiatively recombine with holes supplied by the substrate. The (undoped) AlGaAs top layer reduces the non-radiative surface recombination and the tunneling hole current and thus further improves the optical efficiency.

The experimental measurement technique was introduced earlier[9,10] and exploits spin manipulation in the semiconductor (the Oblique Hanle effect). In brief, under electrical spin injection from the in-plane magnetized FM the injected electrons have an in-plane spin orientation. The application of an oblique magnetic field **B** (in an angle of 45° with the plane) causes a spin precession round **B** with the Larmor frequency ($\Omega = \frac{g^* \cdot \mu_B}{h} \cdot \mathbf{B}$, where $g^*$ is the effective g-factor and $\mu_B$ the Bohr magneton) inside of the semiconductor. This precession leads the average electron spin to acquire a non-zero perpendicular component. This technique allows direct measurements of the steady-state spin polarization *and* the spin lifetime in the semiconductor in the same experiment. However, the application of the oblique magnetic field will slightly tilt the magnetization of the thin film out of the plane. In such a geometry, the circular polarization of the emitted light as a function of the applied oblique magnetic field can be calculated from elementary Bloch equations[13] to be

$$P(B) = \frac{\Pi_{inj} \cdot \cos(\theta)}{4} \cdot \frac{T_S}{\tau} \cdot \frac{(\Omega \cdot T_S)^2}{1 + (\Omega \cdot T_S)^2} + \frac{\Pi_{inj} \cdot \sin(\theta)}{2} \cdot \frac{T_S}{\tau} \cdot \frac{1 + 1/2 \cdot (\Omega \cdot T_S)^2}{1 + (\Omega \cdot T_S)^2} \quad (1)$$

where *P(B)* is the detected degree of circular polarization, $P = (I^+ - I^-)/(I^+ + I^-)$, $I^+$ and $I^-$ are the intensities of right and left circularly polarized radiation. $\Pi_{inj}$ is the degree of injected spin polarization $\Pi = (n^\uparrow - n^\downarrow)/(n^\uparrow + n^\downarrow)$, $T_S$ is the spin lifetime ($T_S^{-1} = \tau^{-1} + \tau_s^{-1}$, $\tau$ is the electron lifetime, i.e., electron-hole recombination time and $\tau_s$ is the spin relaxation time) and $\theta$ is the angle between the magnetization vector and the plane. The equation is a superposition of two terms, which describe the different contribution of the spin injection caused by the in-plane (left term) and the out-of-plane (right term) component of the magnetization of the thin film.

In our structure, $\theta$ shows an approximately linear behaviour as a function of the applied magnetic field and reaches about 10° at 0.6T (A linear behaviour was observed in AGFM and MCD measurements in an oblique magnetic field.

Extraordinary Hall Effect (EHE) measurements have shown that the out-of-plane magnetization of the thin film saturates at $B_\perp$=1.3 T.

Fig. 2(a) (squares) shows the degree of circular polarization of the emitted light in our devices, which is induced by spin injection, as a function of the oblique magnetic field at 80 K. The curve is strongly non-linear, characteristic for the OHE effect. The contribution of the Magnetic Circular Dichroism (MCD) appearing in the ferromagnetic film and caused by the out-of-plane component of the magnetization of the thin magnetic film is measured in a completely optical experiment[9,10)] and has been subtracted from the measured signal.

The fit using eq.(1) gives a spin polarization $\Pi_{inj} \cdot T_S / \tau = (24 \pm 3)$ % and a spin lifetime $T_s = (140 \pm 10)$ ps. The actual spin polarization of electrons that traversed the ferromagnetic metal / semiconductor interface $\Pi_{inj} = \tau / T_S \cdot 24$ %, is higher than 24 % by the factor $\tau / T_S = (\tau_s + \tau) / \tau_S > 1$. The parameter $T_S / \tau$ describes the spin scattering of electrons during their lifetime on the bottom of the conduction band in the semiconductor. The $T_S / \tau$ ratio is not known for the active region of the device since in the supporting optical injection-detection [9,10)] experiment the PL is dominated by the highly doped GaAs substrate.

The room temperature results are presented in Fig. 2b after subtraction of the MCD-contribution. The measured curve now looks linear, as expected, since at room temperature the enhanced spin scattering leads to a broadening of the Hanle curve. Within the small range of the magnetic field this does not allow observation of saturation of the measured circular polarization. The most conservative Hanle fit reveals a spin polarization of $\Pi_{inj} \cdot T_S / \tau =(4.7 \pm 1)$ % and a spin lifetime $T_s$ of $(55 \pm 20)$ ps.

The difference in $\Pi_{inj} \cdot T_S / \tau$ and $T_S$ at 80K and RT is not surprising and entirely reflect the spin properties within the GaAs spin detector. It is well known[14] that the electron lifetime $\tau$ slightly increases with temperature and that the spin relaxation time $\tau_S$ rapidly decreases with temperature. As a result one can clearly observe experimentally the reduction of the spin lifetime $T_S$: $(T_S)_{80K} / (T_S)_{300K} \approx 2.54$. The decrease of the spin lifetime $T_S$, as well as the increase of the electron lifetime $\tau$ both lead to a smaller value of $T_S / \tau$ at room temperature. In this case, assuming $T_S / \tau \sim 1$ at 80 K, the lower bound of polarization of electrically injected electrons is $\Pi_{inj} = 12 \pm 3$ % at RT.

In Fig. 3, the bias dependence of the measured spin polarization of a typical device is shown. It can be clearly seen that the spin polarization decreases for a higher applied bias voltage. We attribute this decrease to an enhanced loss of spin polarization for injected hot electrons during their thermalization in the undoped GaAs active region. This will be discussed elsewhere.

In conclusion, we have shown large spin injection efficiencies for spin injection in a MIS-LED structure at low and at room temperature with an optimized tunnel barrier. The inherent properties of the ferromagnet/AlO$_X$/semiconductor interface (thermodynamical stability, no lattice mismatch problems), the robustness of the fabrication process and the use of the well-established magnetic tunnel junction (MTJ) technology demonstrate the potential of MIS injectors for other spintronic devices.

We thank Willem van de Graaf and Stefan Degroote for MBE sample growth, Barun Dutta for discussions and Olivier Richard for the TEM. PVD and JD acknowledge financial support from the I.W.T. (Belgium) and WVR from the F.W.O.

(Belgium). This work is supported in part by the EC project SPINOSA (IST-2001-33334) and as IMEC Innovation Project.

FIGURE CAPTIONS

**Fig. 1:** Transmission Electron Microscopy (TEM) picture of the CoFe-AlO$_X$-AlGaAs interface of the fabricated spin-LED. In the inset a magnification of the interface is shown.

**Fig. 2:** Degree of circular polarization as a function of the oblique magnetic field in the MIS spin-LED at 80 K (a) and 300 K (b) (The signal has been corrected for the MCD-effect) and the OHE (eq. (1)) fit (solid line) with following fit parameters: Spin polarization in the semiconductor at 80K: $\Pi_{inj} \cdot T_S / \tau = 24 \pm 1\%$, at 300K: $\Pi_{inj} \cdot T_S / \tau = 4.7 \pm 1\%$ and spin lifetime at 80K: $T_s = 0.14 \pm 0.01$ns, at 300K: $T_s = 55 \pm 20$ps.

**Fig. 3:** The injected spin polarization in the GaAs active region as a function of the bias voltage for 300K (■) and 80 K (●). The inset shows two Hanle curves for two different bias points at 80 K and 300 K.

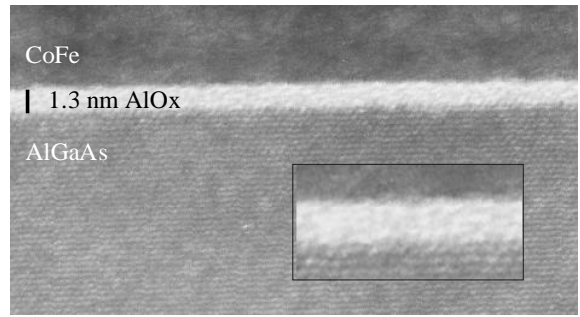

**Fig. 1 (width : 8,5 cm) -- Van Dorpe *et al.***

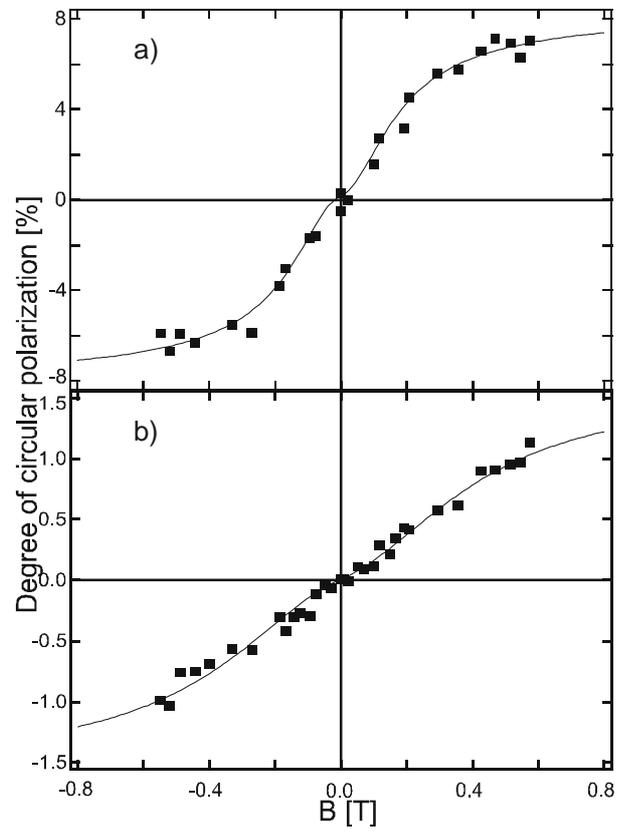

**Fig. 2 (width=8,5 cm) -- Van Dorpe *et al.***

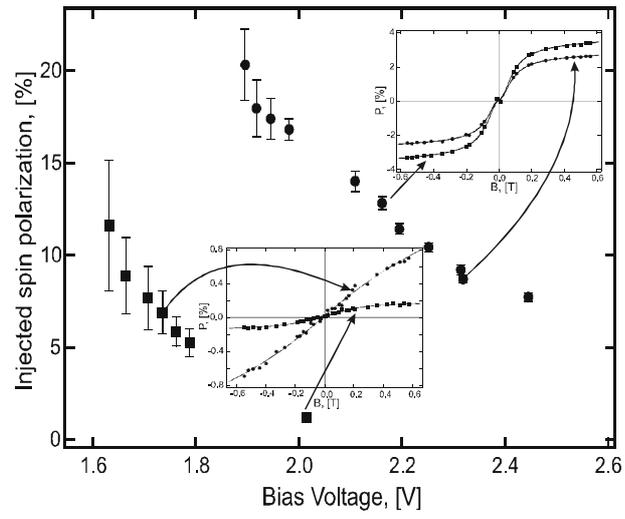

**Fig. 3 (width=8,5 cm) – Van Dorpe _et al._**